\documentclass[%
reprint,
superscriptaddress,
nobibnotes,
secnumarabic,
amsmath,amssymb,
prb,
]{revtex4-1}

\usepackage{amsmath}
\usepackage{graphicx}
\usepackage{float}
\usepackage{bm}
\usepackage[utf8]{inputenc} 
\usepackage{hyperref}
\usepackage{xcolor}

\newcommand{\angstrom}{\text{\normalfont\AA}}
\makeatletter
\newcommand*{\rom}[1]{\expandafter\@slowromancap\romannumeral #1@}
\makeatother
\begin{document}


\title{Commensurate and incommensurate double moir\'e interference in twisted trilayer graphene}

\author{Hai Meng}
\affiliation{Key Laboratory of Artificial Micro- and Nano-structures of Ministry of Education and School of Physics and Technology, Wuhan University, Wuhan 430072, China}

\author{Zhen Zhan}
\email{Corresponding author: zhen.zhan@whu.edu.cn}
\affiliation{Key Laboratory of Artificial Micro- and Nano-structures of Ministry of Education and School of Physics and Technology, Wuhan University, Wuhan 430072, China}



\author{Shengjun Yuan}
\email{Corresponding author: s.yuan@whu.edu.cn}
\affiliation{Key Laboratory of Artificial Micro- and Nano-structures of Ministry of Education and School of Physics and Technology, Wuhan University, Wuhan 430072, China}
\affiliation{Wuhan Institute of Quantum Technology, Wuhan 430206, China}

\date{\today}

\begin{abstract}

Twisted graphene multi-layers have been recently demonstrated to share several correlation-driven behaviours with twisted bilayer graphene. In general, the van Hove singularities (VHSs) can be used as a proxy of the tendency for correlated behaviours. In this paper, we adopt an atomistic method by combining tight-binding method with the semi-classical molecular dynamics to investigate the electronic structures of twisted trilayer graphene (TTG) with two independent twist angles. The two independent twist angles can lead to the interference of the moir\'e patterns forming a variety of commensurate/incommensurate complex supermoir\'e patterns. In particular, the lattice relaxation, twist angle and angle disorder effects on the VHS are discussed. 
We find that the lattice relaxation significantly influence the position and magnitude of the VHSs. In the supermoir\'e TTG, the moir\'e interference provides constructive or destructive effects depending on the relative twist angle. By modulating the two independent twist angles, novel superstructures, for instance, the Kagome-like lattice, could constructed via the moir\'e pattern. Moreover, we demonstrate that a slight change in twist angles (angle disorder) provides a significant suppression of the peak of the VHSs. Apart from the moir\'e length, the evolution of the VHSs and the LDOS mapping in real space could be used to identify the twist angles in the complicated TTG. In practice, our work could provide a guide for exploring the flat band behaviours in the supermoir\'e TTG experimentally. 

\end{abstract}


\maketitle

\section{introduction}
 When stacking two or more two-dimensional van der Waals materials with a lattice mismatch or a relative twist angle, a moir\'e superlatice is formed\cite{geim2013van}. For some moir\'e superlattices, a distinguishing feature is the appearance of flat bands at charge neutrality for which the renormalized Fermi velocity is zero, resulting in the Coulomb interaction strength to significantly exceed the kinetic energy of electrons in the flat band, favouring electron-electron correlations\cite{bistritzer2011moire}. Flat bands appear in twisted bilayer graphene (TBG) with the twist angle equals 1.05$^\circ$, termed the first magic angle, which exhibits a very interesting range of exotic phenomena including Mott insulating\cite{cao2018correlated}, superconductivity\cite{cao2018unconventional}, ferromagnetism\cite{sharpe2019emergent}, Chern insulators\cite{nuckolls2020strongly}, quantum anomalous Hall effect (QAHE)\cite{serlin2020intrinsic} and ferroelectricity\cite{klein2022electrical}. These exciting discoveries have inspired a vast theoretical and experimental search to extend the family of moir\'e superlattices that exhibit the correlation-driven behaviours, including twisted monolayer-bilayer graphene\cite{polshyn2020electrical,xu2021tunable}, twisted bilayer-bilayer graphene\cite{cao2020tunable,he2021symmetry}, trilayer graphene on hexagonal boron nitride\cite{chen2019evidence,chen2019signatures}, twisted multilayer graphene\cite{park2022robust,burg2022emergence} and transition metal dichalcogenides\cite{tang2020simulation,wang2020correlated}. These moir\'e superlattices share both similarities and differences with the TBG in symmetries, band topology and interaction strength, which could help us have a deeper understanding of the correlated behaviours in TBG. Compared to the TBG, some of the moir\'e superlattices may have practical advantages in fabrication or tunability of the physical properties.       


Recently, twisted trilayer graphene (TTG) has gained extensive attention due to the presence of unconventional correlated states\cite{park2021tunable, cao2021pauli}.  
In twisted trilayer graphene, new tunable degrees of freedom are introduced by the addition of an extra third layer on the bilayer graphene. For instance, in the TTG with two consecutive twist angles $\theta_{12}$ and $\theta_{23}$, the beatings of two bilayer moir\'e patterns may lead to a more complex supermoir\'e pattern\cite{zhu2020twisted,zhang2021correlated}. In fact, the experimental technique to realize the TTG is readily available\cite{}. Different from the TBG, electronic structures of TTG are highly dependent on the original stacking arrangements and on which layer is twisted\cite{polshyn2020electrical,wu2021lattice}, and are more sensitive to external perturbations\cite{wu2021magic,lopez2020electrical,chen2021electrically}. In mirror symmetric TTG, a set of dispersive bands coexists with flat bands at charge neutrality, and the magic angle is $\sqrt{2}$ times larger than that of the TBG\cite{carr2020ultraheavy}. Robust and highly tunable superconductivity has been observed in magic angle mirror symmetric TTG\cite{park2021tunable, cao2021pauli}. It has been reported that TTG without mirror symmetry hosts a variety of correlated metallic and insulating states, and topological magnetic states\cite{polshyn2020electrical,chen2021electrically,xu2021tunable}. The magic angle of such low symmetry TTG approximates that of the TBG, and possesses correlated states that are asymmetric with respective to the external electric field\cite{xu2021tunable,ma2021topological}.

\begin{figure}[t]	
	\includegraphics[width=0.45\textwidth]{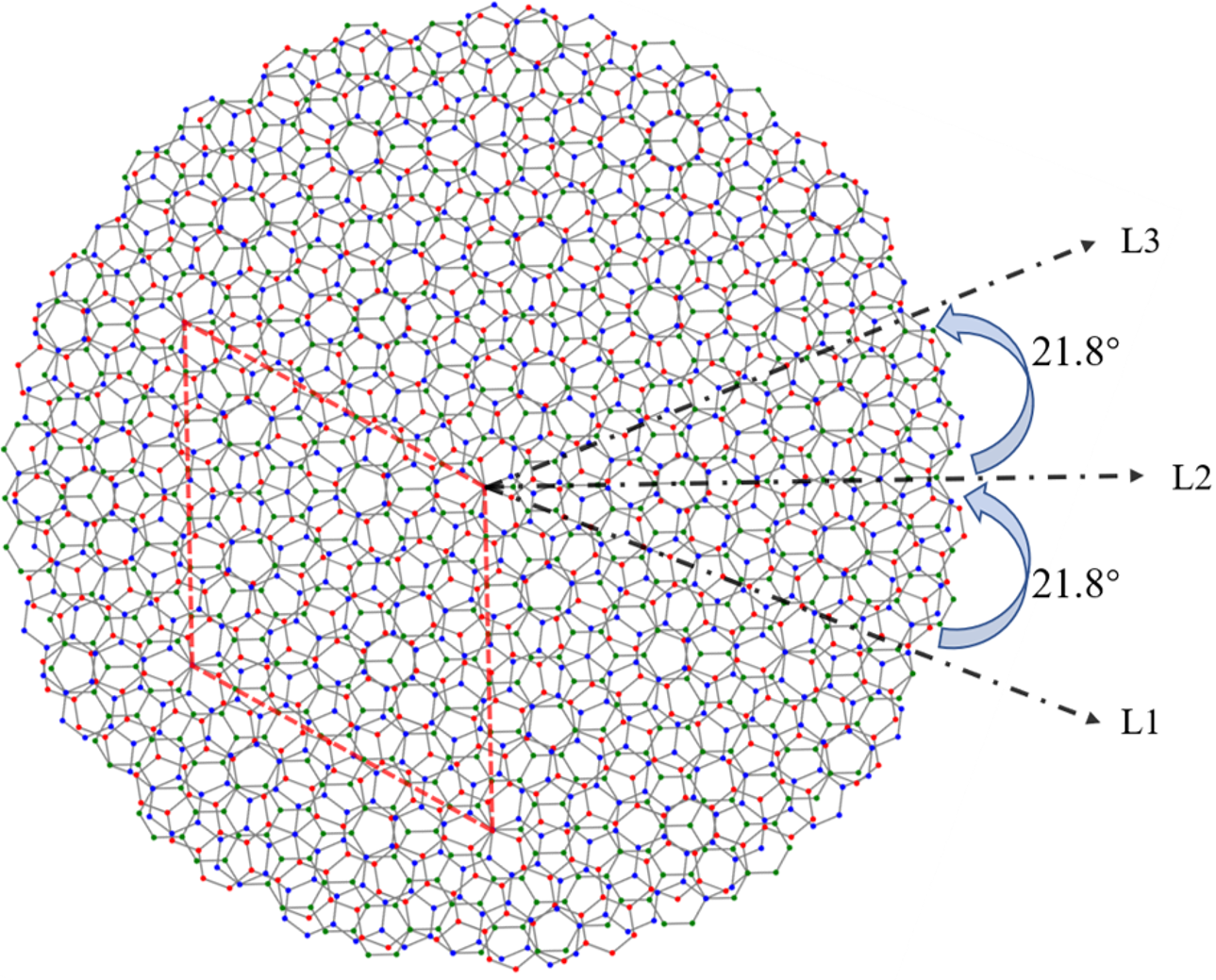}
	\caption{Schematic of the supermoir\'e trilayer graphene with $\theta_{12}=\theta_{23}=21.8^\circ$. The $\theta_{12}$ and $\theta_{23}$ are twist angles between L1 and L2, L2 and L3, respectively. The atoms in L1 (bottom layer), L2 (middle layer) and L3 (top layer) are represented by  red, green and blue dots, respectively. The unit cell of the moir\'e supercell is outlined in a red dashed line.}
	\label{fig:geometry}
\end{figure}

Compared to the TBG, there are many new challenges in the theoretical calculations of the electronic structures of supermoir\'e TTG: the large size of the moir\'e pattern and the lack of commensurate supercell for general twist angles. Firstly, the system size becomes much larger. For example, in the mirror symmetric TTG, the number of atoms in a particular twist angle is half more than that of TBG with the same twist angle. When the twist angles are small, the length of the supermoir\'e period can be extremely large. The case of the mirror-asymmetric TTG with two independent twist angles is even worse. For example, the moir\'e length of the TTG in Fig. \ref{fig:geometry} is 2.6 times larger than that of TBG with the same angle. Secondly, because the two twist angles are independent, the supercell description is no longer valid in some cases since the TTG samples become incommensurate. 

In theoretical calculations, electronic structures of twisted multilayer graphene are commonly described by effective continuum models\cite{2020_prl,li2019electronic,lei2021mirror,carr2020ultraheavy}. These continuum models with momentum space basis are efficient for computation and require no constraints on the twist angles to construct commensurate supercells.  
The continuum results show that the van Hove singularities (VHSs) of TTG are significantly dependent on the tuning parameters, for instance, twist angles\cite{zhu2020twisted} and original stacking arrangements\cite{li2019electronic}. Similar to the TBG, the lattice relaxations significantly influence the electronic behaviours of TTG with tiny twist angles and is an important factor to obtain realistic description of the system\cite{zhan2020large}. Up to now, the lattice relaxation effect is only considered via a continuum model to consider the in-plane distortions and a generalized stacking fault energy to account for the interlayer coupling\cite{carr2020ultraheavy,zhu2020modeling}. An atomistic simulation of the lattice relaxation effects on the electronic structures of supermoir\'e TTG is still missing.   In practice, although the control of the global twist angle with a precision of about 0.1 degrees has been achieved, twist-angle variation across different parts of the device with the order of $\pm 0.01^\circ$ are still exists\cite{uri2020mapping,kazmierczak2021strain}. How will such angle disorder affect the electronic structures of TTG?  

In this paper, we systematically investigate the moir\'e interference in both commensurate and incommensurate TTG. In particularly, the lattice relaxation, twist angle and angle disorder effects on the electronic structures of TTG with two independent twist angles are studied. To address the challenges of the lack of periodicity and the large system size, we adopt a round disk method to construct the TTG samples with arbitrary twist angles and calculate the electronic properties via the tight-binding propagation method (TBPM) implemented in our home-made package TBPLaS\cite{tbplas}. The TBPM is based on the numerical solution of time-dependent Schr\"{o}dinger equation and requires no diagonalization processes. Importantly, both memory and CPU costs scale linearly with the system size. So the TBPM is an efficient method to calculate electronic properties of large-scale and complex quantum systems\cite{yu2019dodecagonal, zhan2020large, kuang_plasmon}. The lattice relaxation is considered via the semi-classical molecular dynamics simulation. We find that the electronic behaviours of TTG are quite sensitive to both the two independent twist angles and angle disorders, in particular for the TTG with mirror symmetry. The systems can have either a constructive or destructive moir\'e interference depending on the two twist angles. Interestingly, by modulating the angles, novel states, for instance the kagome-like states can be constructed in TTG. The VHSs and local density of states (LDOS) mapping can be utilized as quantities to identify the twist angles of TTG in case that the unit cell size corresponds to a range of different possible sets of twist angle pairs.

This paper is organized as follows. In Sec. \ref{sec2}, we introduce the notation and the geometry of the twisted trilayer graphene, the TB model and the computational methods. In Sec. \ref{sec3}, for different twist angle configurations, we show the low energy density of states as a function of twist angle for TTG with and without lattice relaxation. The real space distribution of the electron states of energies at VHS near charge neutral point for different twist angle configurations are also investigated. In Sec. \ref{sec4}, we discuss the effect induced by the twist angle disorder on the VHS. Finally, we give a summary of our work.

\section{Geometry and numerical methods}\label{sec2}

\subsection{Moir\'e structure}
As shown in Fig. \ref{fig:geometry}, we use a round disk method to construct the TTG with arbitrary twist angles. The two independent twist angles $\theta_{12}$ and $\theta_{23}$ are chosen to be the rotation of the second layer L2 relative to the first layer L1 and the rotation of the third layer L3 relative to the second layer L2, respectively. The rotation origin is chosen at an atom site. We use a twist angel pair ($\theta_{12}$, $\theta_{23}$) as the notation for different twist angle configurations. Positive (negative) values of the twist angle denotes counterclockwise (clockwise) rotations. The sample with (-$\theta$, $\theta$) has a mirror symmetry with the middle layer as the mirror plane. Figure \ref{fig:geometry} shows the ($21.8^\circ$, $21.8^\circ$) configuration of twisted trilayer graphene.  

For twisted bilayer honeycomb structures, grahene on hexagonal boron nitride (hBN) for instance, an expression for the period of moir\'e pattern is given by:
\begin{equation}\label{m_period}
\lambda = a\frac{1+\delta}{\sqrt{2(1+\delta)(1-\cos{\theta})+\delta^2}}
\end{equation}
where $a$ is the lattice constant of graphene, $\delta$ is the lattice mismatch between the two two-dimensional (2D) materials. The relative rotation $\phi$ between the moiré pattern and the reference layer is given by:
\begin{equation}\label{t_angle}
\tan{\phi} = \frac{\sin{\theta}}{(1+\delta)-\cos{\theta}}
\end{equation}
For TBG, lattice mismatch $\delta=0$ and Eqs. (\ref{m_period}) and (\ref{t_angle}) can be simplified as:
\begin{align}
\label{tbg_m}
\lambda = \frac{a}{2\sin{(\theta/2)}} \notag \\
\tan{\phi} = \frac{\sin{\theta}}{1-\cos{\theta}}
\end{align}

In twisted trilayer graphene, supermoir\'e patterns arise from the interference between the two bilayer moir\'e patterns. From Eq. (\ref{tbg_m}), the bilayer moire period $\lambda_{12}$ ($\lambda_{23}$) and their relative rotation $\theta_{m}$ are:
\begin{gather}
\lambda_{ij} = \frac{a}{2\sin{\left|\theta_{ij}/2\right|}} \notag \\
\theta_{m} = \left|\theta_{12}+\theta_{23}\right|/2
\end{gather}
If $-\theta_{12} \neq \theta_{23}$, there is a mismatch between these two moir\'e pattern. Then by substituting $\lambda_{12}$, $\lambda_{23}$ and $\theta_{m}$ into Eq. (\ref{m_period}), the trilayer supermoir\'e period can be obtained. When $\theta_{12}\cdot\theta_{23}>0$:
\begin{equation}
\lambda_{MoM} =\frac{a}{2}\left[ \frac{3}{2}+\frac{\cos{(\theta_{12}+\theta_{23})}}{2}-\cos{\theta_{12}}-\cos{\theta_{23}} \right]^{-\frac{1}{2}}
\label{SM_length}
\end{equation}
and when $\theta_{12}\cdot\theta_{23}<0$:
\begin{equation}
\lambda_{MoM} =\frac{a}{2}\left[ \frac{1}{2}-\frac{\cos{(\theta_{12}+\theta_{23})}}{2} \right]^{-\frac{1}{2}}
\end{equation}
The real space period of the supermoir\'e pattern of small angle twisted trilayer graphene is very large in most cases. Moreover, for general twist angle pairs, a commensurate supercell does not exist. To calculate the property of these large scale systems with arbitrary twist angles, we construct the system in a large round disk. The radius of the disk should be set sufficiently large to rid the effects of edge states\cite{yu2019dodecagonal} and to cover the large moir\'e period. In the actual calculation, the disk with radius of $172.2$ nm ($700a$) and contains 10 million carbon atoms are large enough for the twist angles investigated in the paper. 

\subsection{Lattice relaxation}
In the round disk model, carbon atoms at the edge of the disk has dangling bonds which destabilize the system in the process of the relaxation. Thus, the edge carbon atoms are passivated by hydrogen atoms to saturate the dangling $\sigma$ edge bonds. The passivation is implemented by placing in-plane hydrogen atoms for each graphene layer near carbon atoms possessing dangling bond. The carbon-hydrogen bond length is assumed to be 0.1 nm\cite{Subrahmanyam_CH_bond}. 
For the simulation of the structure relaxation, we employ the classical molecular dynamics simulation package LAMMPS\cite{LAMMPS} to do the full lattice relaxation. Intra-layer $C-C$ and $C-H$ interactions are simulated with REBO potential\cite{Rebo_2002}. Inter-layer $C-C$ interaction are simulated with the kolmogorov/crespi/z version of Kolmogorov-Crespi potential\cite{KC}. This method has been used to study the atomic relaxation effect of other twisted multilayer graphene strucures\cite{Wijk_2015, yu2019dodecagonal}.

\subsection{Tight-binding model}\label{sec-2-3}
In this paper, we use a parameterized full tight-binding (TB) scheme for our calculation\cite{full_tb2012numerical}. The form of the Hamiltonian of twisted trilayer graphene (tTG) can be written as
\begin{equation}\label{Hamil}
H = \sum_i\epsilon_i |i \rangle\langle i|+\sum_{\langle i,j\rangle}t_{ij} |i \rangle\langle j|,
\end{equation} 
where $|i\rangle$ is the $p_z$ orbital located at $\mathbf{r}_{i}$, and $\langle i,j\rangle$ is the sum over index $i$ and $j$ with $i \neq j$.  
The hopping integral $t_{ij}$, interaction between two $p_z$ orbitals located at $\mathbf{r}_{i}$ and $\mathbf{r}_{j}$ is\cite{Slater_Koster}
\begin{equation}
t_{ij}=n^2V_{pp\sigma}(r_{ij})+(1-n^2)V_{pp\pi}(r_{ij}),
\end{equation}
where $r_{ij}=|\mathbf{r}_{j}-\mathbf{r}_{i}|$ is the distance between $i$ and $j$ sites, with $n$ as the direction cosine along
the direction $\bm{e_z}$ perpendicular to the graphene layer . The Slater and Koster parameters $V_{pp\pi}$ and $V_{pp\sigma}$ :
\begin{equation}
\begin{aligned}
V_{pp\pi}(r_{ij})=-\gamma_0e^{q_\pi(1-r_{ij}/d)}F_c(r_{ij}),\\
V_{pp\sigma}(r_{ij})=\gamma_1e^{q_\sigma(1-r_{ij}/h)}F_c(r_{ij}),
\end{aligned}
\end{equation}
where $d=1.42\; \angstrom$ and $h=3.349\; \angstrom$ are the nearest in-plane and out-of-plane carbon-carbon distance, respectively, $\gamma_0$ and $\gamma_1$ are commonly reparameterized to fit different experimental results\cite{lammps2019continuum,leconte2022relaxation}. Here we set $\gamma_0=3.2$ eV and $\gamma_1=0.48$ eV. The parameters $q_\sigma$ and $q_\pi$ satisfy $\frac{q_\sigma}{h}=\frac{q_\pi}{d}=2.218 \angstrom^{-1}$, and the smooth function is $F_c(r)=(1+e^{(r-r_c)/l_c})^{-1}$, where $l_c$ and $r_c$ are chosen as $0.265$ and $6.14\; \angstrom$, respectively. We only consider the interlayer hoppings between adjacent layers.

\subsection{The density of states and quasieigenstates}
Each round disk contains more than ten millions of atoms, which is beyond the capability of commonly used density-functional theory and TB based on diagonalization process. We adopt the TBPM to calculate electronic properties of TTG in a round disk\cite{Hans_2000, yuan2010tipsi}.  For the density of states (DOS), the detailed formula is
\begin{equation}\label{dos}
D(\varepsilon)=\frac{1}{2\pi S}\displaystyle\sum_{p=1}^{S}\int_{-\infty}^{\infty}e^{i\varepsilon t}\langle\varphi_p(0)|e^{-iHt}|\varphi_p(0)\rangle dt,
\end{equation}
where $|\varphi_p(0)\rangle$ is one initial state which is the random superposition of all basis states, 
$S$ is the number of random initial states. The calculation error vanishes with $\sqrt{SN}$\cite{Hans_2000}. $N$ is the dimension of the Hamiltonian matrix which equals to the number of atoms in the graphene tight-binding model.  In the round disk TTG with radius of $172$ nm, the number of atoms is around 10 million. Thus, in real calculation, a relatively small finite value of $S$ is sufficient to obtain a convergent results. We use the same simulation parameters in all the calculations.

\begin{figure*}[t!]
	\includegraphics[width=0.9\textwidth]{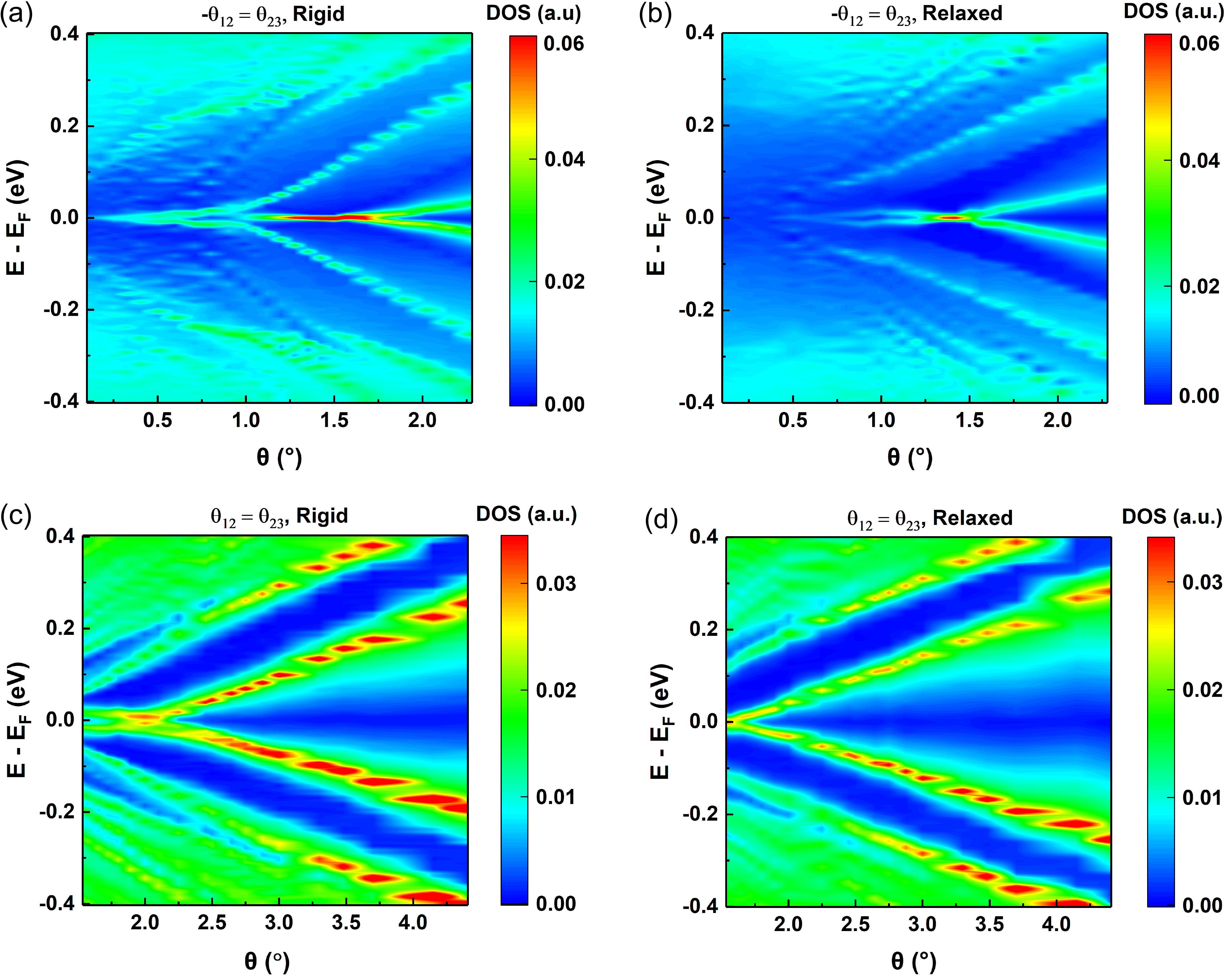}
	\caption{The density of states (DOS) as a function of twist angles varying from $0.1^\circ$ to $2.28^\circ$ for $-\theta_{12}=\theta_{23}$ in (a) rigid and (b) relaxed twisted trilayer graphene. The DOS as a function of twist angles varying from $1.35^\circ$ to $4.4^\circ$ for $\theta_{12}=\theta_{23}$ in (c) rigid and (d) relaxed twisted trilayer graphene. }
	\label{fig:rigidABC}
\end{figure*}

The distribution of states in real space can be obtained by calculating the quasieigenstates\cite{yuan2010tipsi} (a superposition of degenerate eigenstates with certain energy). The quasieigenstates has the expression:
\begin{equation}\label{quasi}
|\Psi(\varepsilon)\rangle=\frac{1}{\sqrt{\sum_n|A_n|^2\delta(\varepsilon-E_n)}}\sum_nA_n\delta(\varepsilon-E_n)|n\rangle,
\end{equation} 
where $A_n$ are random complex numbers with $\sum_{n}|A_n|^2=1$, $E_n$ is the eigenvalue and $|n\rangle$ is the corresponding eigenstate. The local density of states (LDOS) mapping calculated from the quasieigenstates is highly consistent with the experimentally scanning tunneling microscopy $\mathrm dI/\mathrm dV$ mapping\cite{zhan2020large}. 

\section{Tuning the electronic states by twist angles}\label{sec3}
\begin{figure*}[!htbp]
	\includegraphics[width=0.9\textwidth]{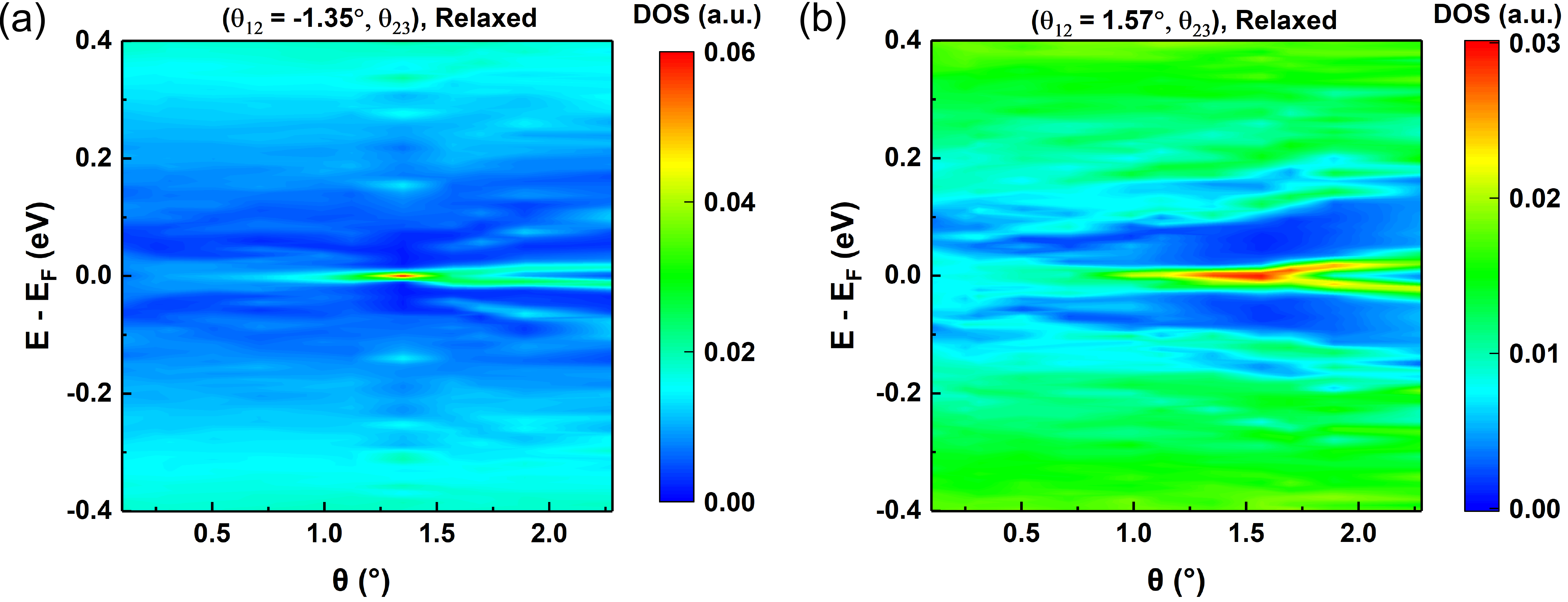}
	\caption{(a) The DOS as a function of $\theta_{23}$ and $\theta_{12}=-1.35^\circ$ in case one of TTG. (b) The DOS as a function of $\theta_{23}$ and $\theta_{12}=1.57^\circ$ in case two of TTG.}
	\label{fig:rigid1.57}
\end{figure*}

In this part, we investigate the twist angle and lattice relaxation effects on the electronic properties of TTG. As we mentioned before, the structures of TTG are strongly dependent on the original stacking arrangements and on which layer is twisted. Here we mainly focus on two different cases: the case \uppercase\expandafter{\romannumeral1} is the samples with $-\theta_{12}=\theta_{23}$, which have mirror symmetry; the case \uppercase\expandafter{\romannumeral2} is a stack of graphene layers where each layer is rotated by a constant amount with respect to the previous one. Each sample in case \uppercase\expandafter{\romannumeral2} has two consecutive twist angles with $\theta_{12}=\theta_{23}$. In case \uppercase\expandafter{\romannumeral1}, the moir\'e length of the TTG is equal to that of TBG with the same twist angle, whereas the moir\'e length of TTG in case \uppercase\expandafter{\romannumeral2} is much larger than both of the bilayer moir\'e lengths due to the interference between the two bilayer moir\'e patterns. According to the Eq. (\ref{SM_length}), if $\theta_{12}=\theta_{23}=4.4^\circ$, the moir\'e length of the TTG is around 41.7 nm. For case \uppercase\expandafter{\romannumeral1} with $-\theta_{12}=\theta_{23}=4.4^\circ$, the moir\'e length is  only around 3.2 nm. The density of states as a function of twist angle $\theta$ to explore the evolution of the van Hove singularities and the lattice relaxation effects are calculated. Due to the incommensurate feature in both cases, we use the round disk model with radius of 172 nm in all calculations. Previous results have been shown that the radius of 172 nm is large enough to get rid of the influence of the edge states\cite{yu2019dodecagonal}. In the round disk method, commensurate or incommensurate systems with any twist angles can be constructed. Therefore, an open boundary is adopted in the calculations.

We first discuss the common features emerging in these two different cases.  As shown in Fig. \ref{fig:rigidABC}, the twist angles significantly modulate the energy position of VHS. The red regions represent VHSs. As the twist angle decreases, the VHS gap decreases first to reach a minimum where the magic angle appears, and then increases in some cases. Moreover, the lattice relaxation obviously modifies the DOS of TTG with tiny twist angles.
For samples in case one without lattice relaxation (rigid), shown in Fig. \ref{fig:rigidABC}(a), two VHSs near charge neutral point (CNP) have their gap narrowing as twist angle $\theta$ decreases, and merge at CNP when angle $\theta=1.57^\circ$ where the DOS reaches the maximum magnitude. We name this angle the magic angle. In fact, a sharp DOS peak located at the CNP appears when $1.2^\circ \leq \theta \leq 1.6^\circ$. With the angle continually decreasing, the VHS gap first increases and then decreases with VHSs merge again at $\theta=0.3^\circ$.  When consider the lattice relaxation (relaxed) in case one, shown in Fig. \ref{fig:rigidABC}(b), the evolution of the VHS gap shows similar tendency as the rigid case. The VHSs merge at CNP and reaches maximum (red area) at the magic angle $1.35^\circ$. The VHS then splits and diminishes as $\theta$ continues to decrease. Compared to the rigid case, the relaxed samples exhibit a narrower range of magic angles, and have VHSs with reduced magnitudes in case of tiny twist angles.     

Compared with ($-\theta$,$\theta$) structures, VHSs evolve differently in ($\theta$,$\theta$) structures as shown in Fig. \ref{fig:rigidABC}(c) and (d). The DOS in case two is orders of magnitude lower than that of the case one. For samples in case two, adjacent layers form two identical bilayer moir\'e periods with a relative rotation of $\theta$ forming a supermoir\'e.  The length of the supermoir\'e period inversely proportional to $1/\theta^2$. For the rigid structures shown in  Fig. \ref{fig:rigidABC}(c), as $\theta$ decreases, the VHS gap narrows and reaches minimum ($20$ meV) at $2.1^\circ$ but the two VHSs never merge at CNP. This result shows agreement with previous study with a continuum model approach\cite{2020_prl}. For the relaxed structures in Fig. \ref{fig:rigidABC}(d), the VHSs merge at CNP at $1.57^\circ$, which is different from the rigid case. 

\begin{figure*}
 	\includegraphics[width=\textwidth]{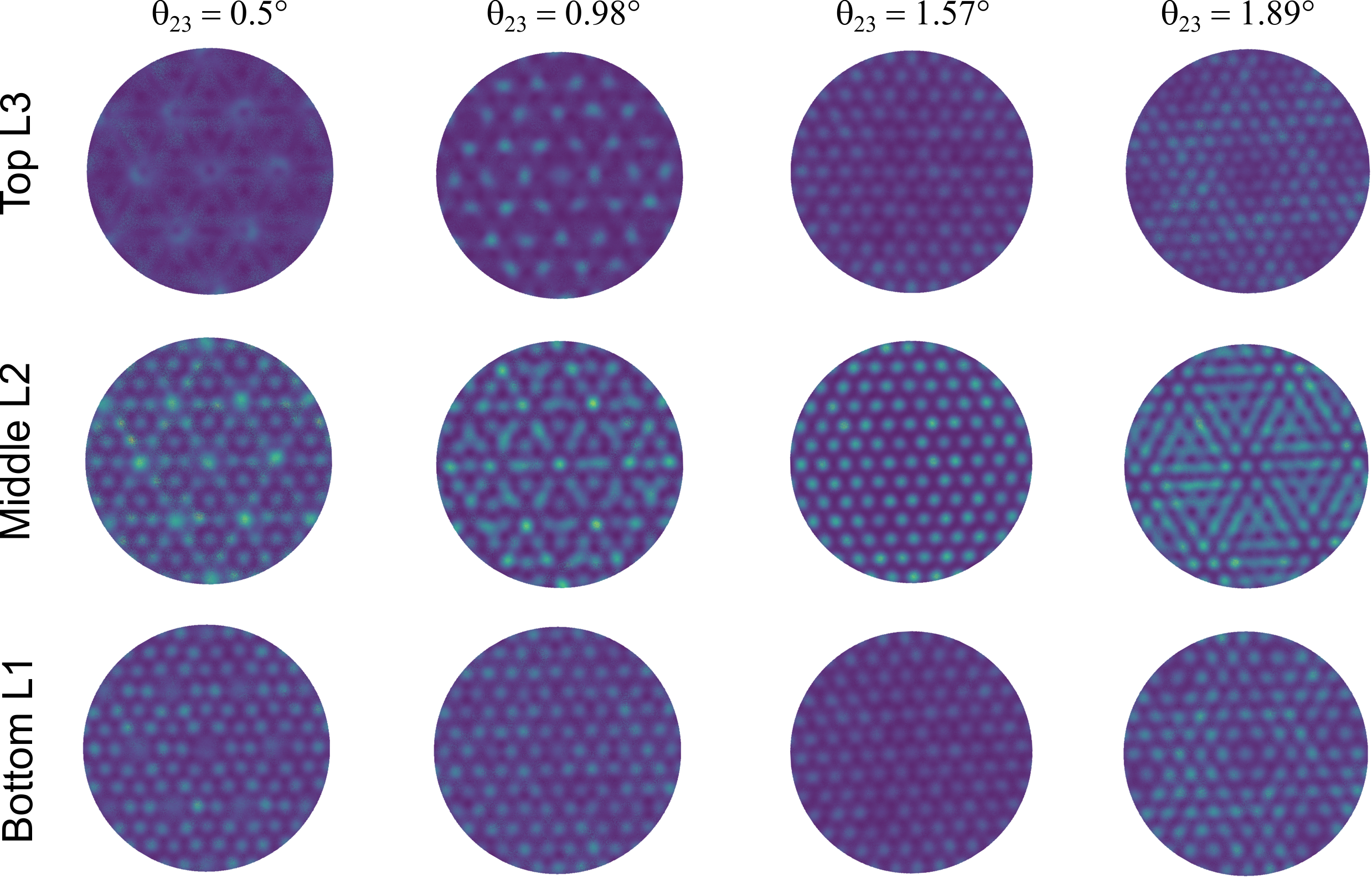}
	\caption{The calculated LDOS mapping of relaxed structures in case two with fifferent $\theta_{23}$ and fixed $\theta_{12}=1.57^\circ$. Energies are selected at VHS in the vicinity of CNP. Brighter area indicates larger density.}
    \label{relax1.57_A}
\end{figure*}

For general twist angles, Fig. \ref{fig:rigid1.57} shows the DOS of relaxed samples as a function of $\theta_{23}$ and fixed $\theta_{12}$. In case one with $\theta_{12}=-1.35^\circ$, when $\theta_{23}=1.8^\circ$, the VHS gap has a value of 30 meV. With $\theta_{23}$ decreasing to the magic angle $1.35^\circ$, the first and second VHSs approach the CNP and the first VHS merge at the CNP when $\theta_{23}=1.35^\circ$. When the $\theta_{23}$ continually decreases, a sharp DOS peak still appears at the CNP but with a lower magnitude due to the destructive interference effects from the superlattice between 2-3 layer pair. For the sample with $\theta_{23}=-\theta_{12}$ where the top and bottom layers are perfectly aligned, there is a strong increases in the DOS due to the constructive interference effects. If we use the presence of VHS as a proxy for electronic correlations, such correlation-driven behaviour is extremely sensitive to a slight change in twist angles, which is similar to previous results\cite{zhang2021correlated}.  
Figure \ref{fig:rigid1.57}(b) shows DOS of relaxed structures in case two with different $\theta_{23}$ and fixed $\theta_{12}=1.57^\circ$. As $\theta$ increases from magic angle $1.57^\circ$, VHSs split with the gap becomes wider, and the magnitude of the VHS becomes smaller. As $\theta$ decreases from $1.57^\circ$, the VHS has a decrease of its magnitude and vanishes as $\theta$ approaches $0^\circ$. Similar to the case one, for $\theta_{23}$ equals to the magic angle, the VHS suffers a constructive interference effect from these two bilayer moir\'e patterns. For $\theta_{23}$ away from the magic angle, the electronic structures can be understood as a magic angle TBG with a destructive interference from the second bilayer moir\'e structure. However, the TTG in case two is less sensitive to the angle disorder than the case one.  For TBG and TTG with mirror symmetry, the twist angle can be identify from the moir\'e length since the unit cell size corresponds to a unique set of twist angles. On the contrary, for TTG without mirror symmetry, a given unit cell size corresponds to a range of different possible sets of twist angle pairs. In this case, the twist angle can be identified by the above VHS evolution with twist angles in experiment. 

 The LDOS mapping is another quantity that can be used to identify the twist angles in experiment. 
 The calculated LDOS mapping of relaxed structures in case two with ($1.57^\circ$,$\theta$), as shown in Fig. \ref{relax1.57_A}, demonstrate the effect of the periodic potential of the moir\'e pattern on the localization of states in real space. We focus on the energies at the van Hove peaks near the charge neutral point in the DOS and show the mapping of the three graphene layers separately. In TTG, the length of the bilayer moir\'e period formed by adjacent graphene layers is $a/2\sin{(\theta_{ij}/2)}$. Smaller relative twist angle gives larger moir\'e  period. In the configuration of ($1.57^\circ$,$\theta$), the bottom and middle layers with twist angle of $1.57^\circ$ forms a moir\'e period of length $\lambda_1=8.98$ nm, the middle and top layers with twist angle of $\theta_{23}$ forms a moir\'e period of length $\lambda_2=a/2\sin{(\theta_{23}/2)}$. Since the ratio of the two moir\'e lengths determines the features of the LDOS mapping, we categorize the mapping results into four types.

\begin{figure*}[htb]
	\includegraphics[width=0.9\textwidth]{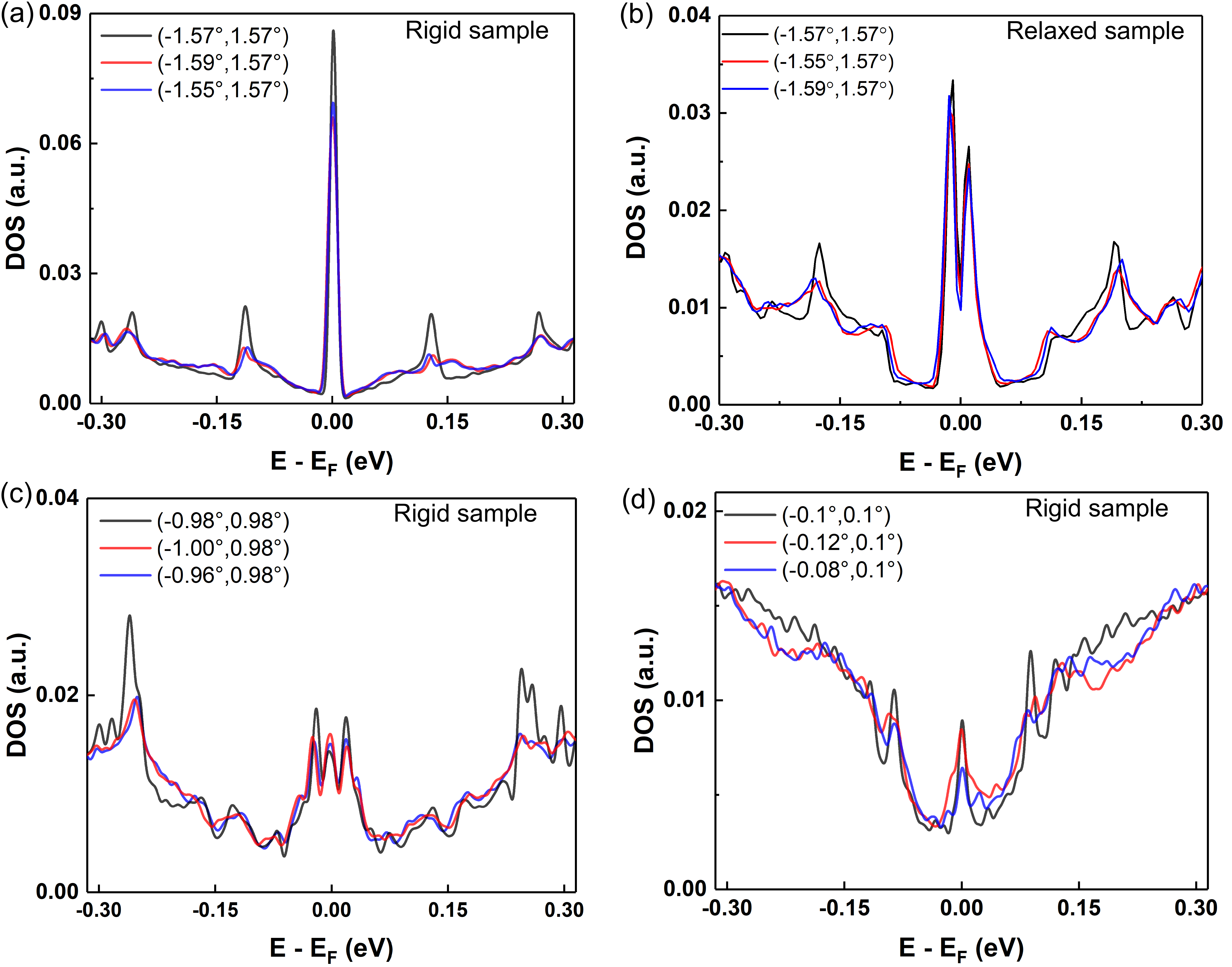}
	\caption{Comparison of the DOS of TTG with (-$\theta_{12}$,$\theta_{23}$) (black line) and with $\theta_{12}$ increased (red line) or decreased (blue line) by $0.02^\circ$ angle disorder. The angle disorder effects in (a) rigid TTG with $-\theta_{12}=\theta_{23}=1.57^\circ$, (b) relaxed TTG with $-\theta_{12}=\theta_{23}=1.57^\circ$, (c) rigid TTG with $-\theta_{12}=\theta_{23}=0.98^\circ$, and (d) rigid TTG with $-\theta_{12}=\theta_{23}=0.1^\circ$.}
	\label{fig.compare}
\end{figure*}

For type $A$ illustrated in Fig. \ref{relax1.57_A} with $\theta_{23}=0.5^\circ$, $\lambda_2$ is much larger than $\lambda_1$, the VHS states are mainly localized on the middle and bottom layers. It is obvious that in the limit $\theta_{12}\gg \theta_{23}$, TTG decomposes into a decoupled TBG moir\'e supercell and a graphene monolayer. The top layer does not contribute significantly to the electronic structures in the low-energy range. That is, the TTG can be considered as a magic angle TBG modified by an effective potential from the third layer.  
In the middle layers, features of both moir\'e patterns are shown, and the localization is enhanced where the $AA$ areas of both moir\'e patterns coincide.
For type $B$ with the ratio of the moir\'e length $\lambda_2/\lambda_1$ around 1.6, for instance $\theta_{23}=0.98^\circ$, the bottom and top layers show features of bilayer (2L) moir\'e pattern slightly affected by the other outer layer, the middle layer gives a new complex supermoir\'e pattern with period $\lambda_3$ larger than both 2L moir\'e periods. 
For type $C$ with $\theta_{23}=1.89^\circ$, the mapping features are similar to that of type $B$. The middle layers also clearly show supermoir\'e patterns much larger than both 2L moir\'e patterns.
For type $D$ with $\theta_{23}=\theta_{12}$ the moir\'e length ratio value $\lambda_1/\lambda_2$ is 1. 
The LDOS mappings only show bilayer magic angle features, and the VHS states are mainly localized in the middle layer. 
Obviously, the LDOS mapping is highly dependent on twist angles. 
By tuning the twist angles, new novel superstructures could constructed. For instance, in case two with $\theta_{12}=1.57^\circ$ and $\theta_{23}=0.98^\circ$, a Kagome-like lattice based on moir\'e pattern are constructed on the top layer. All in all, for the case $\theta_{12} \approx \theta_{23}$, the systems suffer a strong moir\'e interference, and the VHS states are mainly localized in the middle layer. For $\theta_{12} \gg\theta_{23}$, the systems have a weak moir\'e interference, and can be decomposed into a TBG with $\theta_{12}$ and a graphene monolayer.
\section{Twist angle disorder}\label{sec4}

Stacking two 2D materials to an arbitrary angle with a precision of $0.1^\circ$ is still challenging in experiment. Even using the 'tear and stack' technique to fabricate the moir\'e superlattices, twist-angle disorders are still unavoidable from the nonuniformity of the twist angle across the large-scale sample in experiments. In fact, in a high-quality graphene moir\'e pattern, the main source of disorder is the variations of twist angles across the sample. Previous results have been demonstrated that the correlated phases are extremely sensitive to a slight change in twist angles\cite{zhang2021correlated,uri2020mapping}. The angle disorder may explain why two different samples with identical twist angles manifest quite different electronic properties in experiments.  Compared with TBG, the more tunable TTG has more chance suffering the angle disorder since one has to precisely control more than one angle in a sample during the fabrication process. In TTG, the moir\'e length is determined by the two independent twist angles. A small variation of one twist angle could results in a huge increase of the moir\'e length. Then, how will the angle disorder affects the electronic structures of TTG? As we mentioned before, we can construct TTG with arbitrary twist angles by utilizing the round disk method. That is, the twist angles can be continuously tuned no matter whether the system is commensurate or incommensurate. By combining the round disk method with the TBPM,  it is convenient to investigate the angle disorder effects. In this part, we discuss the electronic structures of TTG in the presence of angle disorder. In particular, we focus on the angle disorder effects on the VHSs.    

In TTG structures of case one with ($-\theta$,$\theta$) where the two moir\'e pattern align, we introduce misalignment by an extra twist angle deviation $\Delta\theta=0.02^\circ$ for $\theta_{12}$, i.e. ($-\theta\pm0.02^\circ$,$\theta$). Figure \ref{fig.compare} shows the effect of twist angle disorders on the DOS of TTG with three different angles  $\theta=1.57^\circ,\;0.98^\circ,\;0.1^\circ$ where $\theta=1.57^\circ$ is the magic angle. Let us first focus on the magic angle case. In rigid magic angle TTG, with a variation of $\Delta \theta$ in only $\theta_{12}$, the positions as well as the width of the VHS are unaffected, whereas the peaks of the VHSs are remarkably suppressed. If we quantify the angle disorder effects by a "BCS superconducting transition temperature", which is approximately described as $T_c \propto exp(-\frac{1}{g\rho(E_{vH})})$, our results suggest that the angle disorder strongly suppresses $T_c$. In the $T_c$ expression, $E_{vH}$ is the energy of the VHS peak, $g$ is the electron-phonon coupling. On the contrary, in relaxed magic angle TTG, the positions, width and amplitude of the VHSs are unaffected by the angle disorder. The results in Fig. \ref{fig.compare}(b) shows that the TTG systems with mirror symmetry are protected against twist angle disorders, which is consistent with previous results\cite{carr2020ultraheavy}. In TTG with twist angles smaller than magic angles, the VHS is quite sensitive to the twist angle disorder. For TTG with $\theta=0.98^\circ$, the angle disorders smear some peaks located round 0.3 eV in the DOS. In TTG with $\theta=0.1^\circ$, both the width and the amplitude of the first and second VHS are modulated by the angle disorder.

\section{conclusion}
In summary, we adopt a real-space and atomistic approach for the simulation of the electronic behaviour and the relaxation effect of twisted trilayer graphene for general twist angle pairs. We demonstrate how the position and strength of the VHSs evolve with the twist angle in different situations. We mainly focus on two different structures: one has a mirror symmetry with angle pair (-$\theta$,$\theta$) and the other has two consecutive angles with ($\theta$,$\theta$). Due to the moir\'e interference, the moir\'e length of the systems in case two are far more larger than that of case one with the same angle $\theta$. Our results show that the atomic relaxations have significant effects on the VHS properties and the emergence of magic angle. The position of the VHSs are highly dependent on the twist angle. We find that for mirror symmetric ($-\theta$,$\theta$) structures, the magic angles are $1.57^\circ$ in rigid case and $1.35^\circ$ in relaxed case. For ($\theta$,$\theta$) structures, the magic angles are $2.1^\circ$ and $1.57^\circ$ in the absence and presence of lattice relaxations, respectively. In TTG with two independent twist angles, when one of the twist angles deviates from the magic angle within a range, the VHS evolution shows a destructive moir\'e interference effect from the angle changes, and the VHSs follow only with the magic angle. We then show that the LDOS mappings provide an intuitive real-space representation of the double moir\'e interference and the resulting large complex supermoir\'e pattern. We find that the mapping features are closely related to the ratio of the two moir\'e length. For the case that the two angles are identical, the system suffers a strong moir\'e interference effect and the VHSs states are mainly localized in the middle layer. For one angle far away from the other, the system has a weak interference and can be decoupled into a TBG and a monolayer graphene. By modulating the twist angles, Kagome-like states are constructed due to the interplay between different layers. More importantly, we could use the LDOS and LDOS mapping to identify the twist angles of the TTG in case that the unit cell size corresponds to a range of different possible sets of twist angle pairs.

For mirror symmetric ($-\theta$,$\theta$) TTG that detected a superconductivity, we find that the twist angle disorder which breaks the mirror symmetry strongly affects the VHS property. This may explain why two different samples with identical twist angles manifest different electronic properties in experiments. We also find that the relaxation weakens the disorder effect. In fact, for such small disorders, the relaxation can restore the system to the favorable AA stacking for outer layers.
Our results could provide a guide for the experiment to explore the flat band behaviours in supermoir\'e TTG.


\begin{acknowledgments}
We thank Gudong Yu for providing the code to generate the round disk TBPLaS. This work was supported by the National Natural Science Foundation of China (Grants No. 12174291 and No. 12047543). S.Y. acknowledges funding from the National Key R\&D Program of China (Grant No. 2018YFA0305800). Numerical calculations presented in this paper have been performed on the supercomputing system in the Supercomputing Center of Wuhan University.
\end{acknowledgments}
\bibliographystyle{apsrev4-1}
\bibliography{reference}

\end{document}